\documentclass[fleqn,usenatbib]{mnras}

\usepackage{newtxtext,newtxmath}

\usepackage[T1]{fontenc}
\usepackage{ae,aecompl}
\usepackage{pbox}
\usepackage{xfrac}
\usepackage{mathtools} 
\usepackage{orcidlink}

\usepackage{graphicx}	
\usepackage{amsmath}	
\usepackage{amsfonts}
\usepackage{gensymb}
\usepackage{subfig}
\usepackage{array}
\newcolumntype{P}[1]{>{\centering\arraybackslash}p{#1}}
\newcolumntype{$}{>{\global\let\currentrowstyle\relax}}
\newcolumntype{^}{>{\currentrowstyle}}
\newcommand{\rowstyle}[1]{\gdef\currentrowstyle{#1}%
  #1\ignorespaces
}
\usepackage{todonotes}

\usepackage{etoolbox}


\title[Line-Driven Accretion Disc Winds]{State-of-the-art simulations of line-driven accretion disc winds:\\ realistic radiation-hydrodynamics leads to weaker outflows}

\author[N. Higginbottom \& N Scepi et. al]
{Nick Higginbottom$^{\orcidlink{0000-0001-7560-4747}~1, \dagger}$,
Nicolas Scepi$^{\orcidlink{0000-0003-3909-2486}~1, \dagger}$,
Christian Knigge$^{\orcidlink{0000-0002-1116-2553}~1}$\thanks{E-mail: C.Knigge@soton.ac.uk},
Knox S. Long$^{\orcidlink{0000-0002-4134-864X}~2,3}$,
James H. Matthews$^{\orcidlink{0000-0002-3493-7737}~4}$ 
\newauthor{and Stuart A. Sim$^{\orcidlink{0000-0002-9774-1192}~5}$
}
\\
$^{1}$School of Physics and Astronomy, University of Southampton, Highfield, Southampton, SO17 1BJ, UK\\
$^{2}$Space Telescope Science Institute, 3700 San Martin Drive, Baltimore, MD, 21218, USA\\
$^{3}$Eureka Scientific Inc., 2542 Delmar Avenue, Suite 100, Oakland, CA, 94602-3017, USA\\
$^{4}$Department of Physics, Astrophysics, University of Oxford, Denys Wilkinson Building, Keble Road, Oxford, OX1 3RH, UK\\
$^{5}$School of Mathematics and Physics, Queen's University Belfast, University Road, Belfast BT7 1NN, UK\\
$^\dagger$ First and second author contributed equally
}

\date{\today}

\pubyear{2023}

\begin{document}
\label{firstpage}
\pagerange{\pageref{firstpage}--\pageref{lastpage}}
\maketitle{}

\begin{abstract}
Disc winds are a common feature in accreting astrophysical systems on all scales. In active galactic 
nuclei (AGN) and accreting white dwarfs (AWDs), specifically, radiation pressure mediated by spectral lines is a 
promising mechanism for driving these outflows. Previous hydrodynamical simulations have largely supported 
this idea, but relied on highly approximate treatments of ionization and radiative transfer. Given the sensitivity of line driving to the ionization state and radiation field in the outflow, here we present 
a new method for carrying out 2.5D radiation-hydrodynamic simulations that takes full 
account of the frequency-dependent radiative transfer through the wind, the corresponding 
ionization state and the resulting radiative accelerations. Applying our method to AWDs, 
we find that it is much harder to drive a powerful line-driven outflow when the 
interaction between matter and radiation is treated self-consistently. This conclusion is  
robust to changes in the adopted system parameters. The 
fundamental difficulty is that discs luminous enough to drive such a wind are also 
hot enough to over-ionize it. As a result, the mass-loss rates in our simulations are 
much lower than those found in earlier, more approximate calculations. We also show that 
the ultraviolet spectra produced by our simulations do not match those observed in AWDs. We 
conclude that, unless the over-ionization problem can be mitigated (e.g. by sub-grid clumping or a 
softer-than-expected radiation field), line driving may not be a promising mechanism for powering 
the outflows from AWDs. These conclusions are likely to have significant implications for disc winds in AGN also.
\end{abstract}



\defcitealias{castor_radiation-driven_1975}{CAK}

\section{Introduction}
\label{section:introduction}
Signatures of outflowing gas are present in a wide range of astrophysical settings 
including young stellar objects (YSOs) \citep[e.g.][]{2018A&A...620L...1G, 2019A&A...631A..74M}, 
hot stars \cite[e.g.][]{2006A&A...446..279C,2018A&A...614A..91H}, accreting white dwarfs
(AWDs) \cite[e.g.][]{2012ApJS..199....7F}, X-ray binaries (XRBs) \citep[e.g.][]{2018Natur.554...69T,
2016AN....337..368D} and quasars/active galactic nuclei (AGN) \citep[e.g.][]{2009ApJ...692..758G}.
In some cases, the driving mechanism is clear. For example, the stellar winds in 
hot OB-stars ($10,000~\mathrm{K}  \lesssim T_{\rm eff,OB} \lesssim 50,000~~\mathrm{K}$) are
almost certainly driven by the radiation pressure that is produced when 
UV photons are scattered by resonance lines in the partially ionized stellar atmosphere. 
The theory of these `line-driven' winds has been developed over the course
of several decades, beginning with the seminal paper by \citet[][hereafter referred to as \citetalias{castor_radiation-driven_1975}] {castor_radiation-driven_1975}.

Line-driving has also been suggested as a possible mechanism for powering the disc winds in 
AWDs and AGN, since the disc temperatures in these systems are comparable to $T_{{\rm eff, OB}}$. In XRBs, line-driving is unlikely to be dominant, because the ionization state of the outflows seen in these system is too high.\footnote{Strictly speaking, this statement only applies to the outflowing plasma that produces the blue-shifted X-ray absorption lines observed in many XRBs. The recent discovery of optical \citep[e.g.][]{2019ApJ...879L...4M,2018MNRAS.481.2646M}  and ultraviolet \citep{2022Natur.603...52C} wind signatures in XRBs implies the presence of lower-ionization outflowing material in at least some of these systems as well.} Here, either thermal (where gas is heated to the point where its thermal velocity exceeds the local escape velocity) or magnetic driving are more promising mechanisms \citep{2016ApJ...821L...9M,2017ApJ...836...42H,2020MNRAS.497.4970T}.

Radiation-driven winds arise in the environments of massive luminous objects whenever the outward force associated with the scattering and/or absorption of photons exceeds the inward pull of gravity \citep{1916MNRAS..77...16E}. The famous Eddington limit is associated with the simplest case of a fully ionized, hydrogen-only plasma, in which case the radiative force is solely due to electron scattering. Line driving is an extension of this mechanism to partially ionized media, where interactions between photons and \emph{bound} electrons can dramatically increase the net radiative force. The ratio of this radiative force relative to the electron scattering case is often referred to as the force multiplier, $\mathcal{M}$, a nomenclature first introduced by \citetalias{castor_radiation-driven_1975}. Force multipliers can reach values as high as $\mathcal{M} \simeq 2000$ if the ionization state of the gas is optimal for line driving \citep{1995ApJ...454..410G}.

\defcitealias{2004ApJ...616..688P}{PK}
\defcitealias{proga_radiation-driven_1998}{PSD}
\defcitealias{dyda_effects_2018}{DP}

Hydrodynamic simulations of line-driven {\em disc} winds have been carried out both for AGN-scale systems (\citealt{2000ApJ...543..686P,2004ApJ...616..688P}, \citetalias{2004ApJ...616..688P}; \citealt{nomura_radiation_2016,nomura_line-driven_2017,nomura_line-driven_2020})
and for AWDs  
(\citealt{proga_radiation-driven_1998}, hereafter \citetalias{proga_radiation-driven_1998}; \citealt{pereyra_hydrodynamical_1997,pereyra_hydrodynamic_2000,dyda_non-axisymmetric_2018}; \citealt{dyda_effects_2018}, hereafter \citetalias{dyda_effects_2018}). A crucial aspect of all such simulations is the spatial distribution of the force multiplier throughout the outflow. The force multiplier depends directly on the ionization state of the wind, which is expected to be set by radiative processes. Of course, the radiation field itself is also modified as a result of its interactions with the outflowing material, so matter and radiation are intricately and non-linearly coupled in line-driven flows. 

In previous hydrodynamic simulations of disc winds, these complex interactions between matter and radiation have been simplified significantly. Usually, this has been achieved 
by considering an isothermal outflow whose ionization state is either completely fixed (e.g. \citetalias{proga_radiation-driven_1998}; \citealt{pereyra_hydrodynamical_1997,pereyra_hydrodynamic_2000,dyda_non-axisymmetric_2018}; \citetalias{dyda_effects_2018}; all in the AWD setting) or estimated via an ionization parameter, $\xi$, 
assuming a radiation field modified only by grey attenuation (e.g. \citealt{2000ApJ...543..686P,2004ApJ...616..688P}, \citetalias{2004ApJ...616..688P}; \citealt{nomura_radiation_2016,nomura_line-driven_2017,nomura_line-driven_2020}; all in the AGN setting). With these assumptions, the force multiplier is then estimated  via the standard \citetalias{castor_radiation-driven_1975} $k$-$\alpha$ parameterization,
 \begin{equation}
   \mathcal{M}(\mathfrak{t})=k\mathfrak{t}^{-\alpha}, 
  \label{eq:k-alpha}
\end{equation}
modulo a slight modification first introduced by \citep{owocki_time-dependent_1988} designed to limit the maximum force muliplier to $\mathcal{M}_\mathrm{max}$ (c.f. Equation~4 in \citetalias{proga_radiation-driven_1998}). Here, $\mathfrak{t}$ is the so-called ``optical depth parameter'', which depends on the local velocity gradient of the flow. The power-law index $\alpha$ measures the relative importance of optically thin and 
optically thick lines. Finally, the normalization constant $k$ encodes the overall efficiency of line-driving. 
For their AWD model, \citetalias{proga_radiation-driven_1998} adopted $k=0.2$ and $\alpha=0.6$, based upon typical 
values for winds in OB stars \citep{1995ApJ...454..410G}. For their AGN simulations, \citet{2000ApJ...543..686P} 
retain $\alpha = 0.6$, but use an analytical approximation for $k(\xi)$, based on the calculations of \cite{1990ApJ...365..321S}. The maximum force multiplier is typically set to $\mathcal{M}_\mathrm{max} \simeq 4400$.

All of these simulations generated line-driven winds that were broadly capable of accounting for the observed outflow signatures in these systems, with typical mass-loss rates on the order of $\dot{M}_{\rm wind} \sim 10^{-3} \, \dot{M}_{\rm acc}$). That said, \cite{DP2000} and \cite{proga_resonance_2002} already noted some tension between simulations and observations of disc winds in AWDs, 
with simulations requiring higher-than-expected accretion rates (by a factor of 2-3) to generate enough wind mass-loss and account for the observed strengths of the wind-formed lines.
In any case, it is not certain that the 
assumptions and approximations that had to be made at the time are entirely benign. For example, 
\citet{2010MNRAS.408.1396S} and \citet{2014ApJ...789...19H} carried out detailed radiative transfer calculations 
for the AGN simulation presented by PK and found that line-driving failed in the simulated outflow, 
because multi-dimensional scattering effects caused the wind to become over-ionized
\footnote{We refer to a wind as ``over-ionized'' if the dominant ionization stages of major elements are significantly higher than those required for efficient line-driving.}.
Similarly, the prescription for the force multiplier adopted in these simulations implicitly assumed that -- just as in OB-stars -- the ionization state in AWDs is near 
optimal for line driving. Moreover, in generating synthetic spectral lines for the simulated outflows, the ionic species associated with these relevant transitions 
were assumed to be dominant. Given all this, it has been a long-standing ambition to carry out new radiation-hydrodynamics (RHD) simulations of line-driven disc winds 
in which multi-frequency radiative transfer and ionization are treated in detail. However, the computational challenges posed by this task are significant.

Here we describe our first attempt to overcome
the above challenges and conduct Monte Carlo radiation-hydrodynamics simulations of a disc wind system in which the radiative transfer and force multiplier are calculated realistically and self-consistently. 
We focus on AWDs as our initial application because, relative to AGN, they can be modelled with a smaller spatial dynamic range and a simpler spectral energy distribution (SED).  The remainder 
of this paper is organized as follows:  In Section~2, we present the method we have 
developed to carry out 2.5D, multi-frequency RHD simulations with a full treatment of ionization and line-driving forces. In Section~3, 
we present the results of several simulations, including some that can be directly compared to those presented by PSD. Finally, 
in Section~4, we discuss our results, explore the limitations of our calculations and summarize our conclusions. 

\section{Method}
\label{section:method}

The fundamentals of our coupled Monte Carlo radiative transfer and hydrodynamics method are as outlined by \cite{2020MNRAS.492.5271H}. Briefly, we 
use an operator-splitting approach in which the hydrodynamic equations, 
\begin{gather}
\frac{\partial\rho}{\partial t}+\pmb{\nabla}\cdot(\rho\textbf{v}) = 0,  \\
\label{eq:angular}
\frac{\partial (\rho\textbf{v})}{\partial t}+\pmb{\nabla\cdot}(\rho\textbf{v}\textbf{v}+p \textbf{I}) = -\rho\pmb{\nabla}\Phi +\rho \textbf{g}_\mathrm{rad},
\end{gather}
are solved using the publicly available Godunov-style code \textsc{pluto} 4.4  \citep{2007ApJS..170..228M}. Here $\rho$, $\textbf{v}$, $p$ and $\Phi$ are respectively the density, the velocity vector, the thermal pressure and the gravitational potential. The radiative force, $\textbf{g}_\mathrm{rad}$, is calculated by using our Monte Carlo ionization and radiative transfer code \textsc{python}\footnote{Python is a collaborative open-source project available at \url{https://github.com/agnwinds/python}.} \citep{2002ApJ...579..725L,2013MNRAS.436.1390H,2015MNRAS.450.3331M}. The two aspects of our computational approach -- hydrodynamics and ionization/radiative transfer -- are described in more detail in  \S\ref{section:pluto_method} and \S\ref{section:python_method}, respectively. Note that we use an isothermal equation of state in \textsc{pluto}, so that we do not have to solve an energy equation.

Our overall iterative numerical procedure can be summarized as follows.
After an initialization stage, which we describe below in \S\ref{section:syspars}, we start by running the hydrodynamic calculation for a short amount of simulation time, $\Delta t_\mathrm{RAD}$ (see \S\ref{section:python_method}), to evolve the density and velocity in the wind. Once this is done, we pass the new velocity, density and temperature fields to \textsc{python} to run a Monte Carlo radiation transport simulation and calculate the ionization state of the gas and the direction-dependent flux in each cell. We then use a stand-alone code (see \S\ref{sec:force_mult}) to calculate force multiplier tables using the new ionization state and direction-dependent fluxes coming from \textsc{python}. These force multiplier tables are passed to \textsc{pluto} along with the direction-dependent flux in order to compute $\textbf{g}_\mathrm{rad}$. To close the loop, we run the hydrodynamics simulation for another $\Delta t_\mathrm{RAD}$. We repeat this entire process  many times, until the wind reaches a quasi-steady state.

All of our simulations employ \textsc{pluto}'s standard hydrodynamics module, with linear reconstruction, and use the Harten, Lax, Van Leer (HLL) approximate Riemann Solver. We use $2^{\rm nd}$ order Runge Kutta time-stepping and adopt a Courant-Friedrichs-Levy number of 0.4 in our calculations. All of our \textsc{python} calculations utilize the so-called ``simple atom'' approach described by \cite{2002ApJ...579..725L}, and we adopt solar abundances throughout. In the rest of this section, we provide more details on each step of these RHD simulations.

\subsection{System Parameters and Initial Conditions}
\label{section:syspars}

\newcolumntype{L}[1]{>{\raggedright\arraybackslash}p{#1}}

\begin{table*} 
\begin{tabular}{$m{0.9cm}^m{2.175cm}^P{0.98cm}^P{2.2cm}^P{1.7cm}^P{0.6cm}^P{0.6cm}^P{1.2cm}^P{0.2cm}^P{0.2cm}^P{1.3cm}^P{0.8cm}}
Model & Comments & 
\normalsize ${\dot{M}_{\rm acc}}\atop{[\dot{M}_{\odot}~{\rm yr}^{-1}]}$& 
$\rm{T_{d,{\rm visc}}(R)}$ &
\normalsize $\rm{Force \atop Multiplier}$ & 
$\rm{\dfrac{L_{\rm BL}}{L_{\rm disc}}}$ &
\normalsize $\rm{T_{\rm BL} \atop [10^5~K]}$ &
$\rm{\dfrac{L_{\rm tot}}{L_{\rm Edd}}}$ & 
PSD & DP & 
\normalsize $\dot{M}_{\rm wind} \atop [\rm{M_{\odot}~{\rm yr}^{-1}}]$ & 
\normalsize $\rm{v_r \atop [km~s^{-1}]}$\\
\\[-0.5em]
\hline \\ [-0.7em]
\rowstyle{\bfseries\boldmath\textbf}
\hspace*{-0.15cm}
HK22D   & Fiducial Model         & $\pi\times10^{-8}$& Shakura-Sunyaev       & self-consistent           & $0$ & $0$     & $1\times10^{-3}$  & 3 & B & $4.6\times10^{-14}$ & 1700\\[0.1cm]
HK22Df  & No RT \& ionization    & $\pi\times10^{-8}$ & Shakura-Sunyaev      & fixed $k$ and $\alpha$    & $0$ & $0$     & $1\times10^{-3}$  & 3 & B & $1.6\times10^{-11}$ & 4000\\
HK22C   & Including BL           & $\pi\times10^{-8}$ & Shakura-Sunyaev      & self-consistent           & $1$ & $1.1$   & $2\times10^{-3}$  & 8 & C & \multicolumn{2}{c}{No Outflow} \\
HK22B1   & Low $\dot{M}_{\rm acc}$    & $\pi\times10^{-9}$ & Shakura-Sunyaev      & self-consistent           & $0$ & $0$     & $1\times10^{-4}$  & 1 &   & \multicolumn{2}{c}{No Outflow} \\
HK22B2   & Low $\dot{M}_{\rm acc}$    & $1\times10^{-8}$ & Shakura-Sunyaev      & self-consistent           & $0$ & $0$     & $3\times10^{-4}$  & 2 & A & \multicolumn{2}{c}{No Outflow} \\
HK22F   & High $\dot{M}_{\rm acc}$   & $\pi\times10^{-7}$ & Shakura-Sunyaev      & self-consistent           & $0$ & $0$     & $1\times10^{-2}$  & 5 &   & $1.2\times10^{-12}$ & 1000\\ 
HK22Ds\footnotemark[1]& Soft SED               & $\pi\times10^{-8}$ & fixed~at~$40,000$~K  & self-consistent           & $0$ & $0$     & $1\times10^{-3}$  &   &   & $4.0\times10^{-13}$ & 500\\
\hline \\
\multicolumn{12}{L{1.0\textwidth}}{$^1$ \footnotesize The characteristic velocity and mass-loss rate may be underestimated for this model, because much of the mass loss takes place in the outer disc, \par \hspace*{0.14cm} near the edge of the computational domain.}
\end{tabular}
\caption{Parameters adopted in the simulations and some derived quantities. For each simulation, we provide the model designation 
 (column ``Model''), a brief description of its key features (``Comments''), the accretion rate (``$\dot{M}_{\rm acc}$''), a brief description of the radial disc temperature distribution (``$\rm{T_{d,{\rm visc}}(R)}$''), a brief description of the prescription used for $\rm{\mathcal{M}(\mathfrak{t})}$ (``Force Multiplier''), the ratio of boundary layer to disc luminosity (``$\rm{L_{BL}/L_{disc}}$''), the boundary layer temperature (``$\rm{T_{\rm BL}}$''), the Eddington fraction $\Gamma$ (``$\rm{L_{tot}/L_{Edd}}$''), the equivalent models in \citetalias{proga_radiation-driven_1998} and \citetalias{dyda_effects_2018} (``PSD'' and ``DP''), the wind mass-loss rate (``$\dot{M}_{\rm wind}$'') and the velocity velocity of the fast parts of the wind  (``$v_r$'').}
\label{table:wind_param}
\end{table*}

In all of our models, the accretor is taken to be a WD with mass $M_{\rm WD} = 0.6~M_{\odot}$ and radius $R_{\rm WD}=8.7\times10^8$~cm. We employ a spherical polar coordinate systems and cover one quadrant of the environment close to the WD for our hydrodynamical calculations. Thus our simulation domain extends from $\theta_{\rm min} = 0$ to $\theta_{\rm max} = \pi/2$ in polar angle and from $r_{\rm min}=R_{\rm WD}$ to $r_{\rm max} \simeq 10 \, R_{\rm WD}$ radially. We use a stretched grid with fixed zone size ratios (as in \citetalias{proga_radiation-driven_1998}), with $dr_{k+1}/dr_{k}=1.07$ and $d\theta_{k+1}/d\theta_{k}=0.95$, to ensure that we achieve the finest resolution in the wind launching region, i.e. close to the accretor and close to the disc plane. The number of grid points in the $r$ and $\theta$ directions are $N_r = 128$ and $N_{\theta} = 96$,  respectively.

For our fiducial model, we focus on the simplest astrophysically interesting case: a system containing a luminous accretion disc in which there is no significant radiation associated with the central object. We adopt an accretion rate of $\dot{M}_{\rm acc} = \pi \times 10^{-8}~\mathrm{M_{\odot}~yr^{-1}}$ for this model, which is very much at the high end of the plausible range for ``normal'' wind-driving AWDs, such as nova-like variables \citep{howell}. This choice is intended to maximize the chance of generating a robust outflow; it also allows a direct comparison to identical models in \citetalias{proga_radiation-driven_1998} and \citetalias{dyda_effects_2018}. 

We also present results for several additional models, characterized by (i) the omission of radiative transfer and ionization effects; (ii) the inclusion of a luminous boundary layer (BL); (iii) a 10$\times$ higher accretion rate; (iv) a 3$\times$ and 10$\times$ lower accretion rate; (v) a flat radial disc temperature profile. Information regarding all of these models is provided in Table~\ref{table:wind_param}.

We assume a geometrically thin disc, so all of the disc radiation emanates from the system mid-plane at $\theta = \pi/2$. Since the disc represents the mass reservoir for the outflow, we fix the density at $\theta = \pi/2$ and hold that value constant throughout the simulation. This density, $\rho_0$, should ideally be set to a value that corresponds to the upper parts of the disc atmosphere, just below the wind launching region. For our simulations, it needs to be set high enough to ensure that the critical and sonic points of the outflow are inside the simulation domain, but low enough to prevent the hydrostatic regions within the domain from becoming completely optically thick. We typically set $\rho_0=10^{-9}\rm{g~cm^{-3}}$ to achieve this goal. We checked that multiplying and dividing $\rho_0$ by a factor of 2 did not change the results of our simulations significantly. Hence, it seems that, as long as $\rho_0$ is set to properly capture the sonic point while avoiding a highly optically thick base of the wind, our results are not very sensitive to the choice of $\rho_0$.

The initial configuration of the simulation in \textsc{pluto} corresponds to a hydrostatic disc atmosphere. Thus the velocities in each grid cell are set to the Keplerian velocity at a given distance from the central source, while the densities are initialized to 
\begin{equation}
    \rho(r,\theta)=\rho_0 \,\, \exp\left(\frac{-GM_{*}}{2c_{\rm iso}^2r~\tan\theta}\right).
\end{equation}
Here, $c_{\rm iso}=\sqrt{kT/\mu m_p}$ is the isothermal sound speed and $\mu=0.6$ is the mean molecular weight. We currently keep the temperature fixed at $T = 40,000$~K throughout the simulation (so that $c_{\rm iso} = 24~\mathrm{km~s^{-1}}$). We have verified that this choice of wind temperature does not have a significant effect on our results (see \S\ref{section:python_method} and \S\ref{section:discussion}).

To initialize the radiative force, we set up a hydrostatic disc atmosphere with a very low density of $10^{-20}\rm{g~cm^{-3}}$ at its base. This ensures that the entire domain is optically thin. We then use \textsc{python} to calculate the corresponding ionization state and direction-dependent fluxes and follow the procedure described at the beginning of \S\ref{section:method} to initialize the radiation force in \textsc{pluto} for the first time step.

\subsection{Hydrodynamics}
\label{section:pluto_method}

We have implemented radiative driving in the hydrodynamics by making use of \textsc{pluto}'s built-in support for ``body forces''. These are user-defined external forces that act on the gas in each grid cell, through a $\rho \textbf{g}_\mathrm{rad}$ term on the right-hand side of the momentum conservation equation. More specifically, during each hydrodynamic time-step, $\Delta t_{\rm HD}$, the net radiative acceleration in each cell is estimated as a
the vector sum, 
\begin{equation}
\textbf{g}_\mathrm{rad}=\sum_i^{N_{\hat{n}}} g_i = \sum_i^{N_{\hat{n}}}{\left(1+\mathcal{M}\left(\mathfrak{t}_i\right)\right)\,\,\sigma_e\,\,\frac{\vec{F}_{{\rm UV},i}}{c}}.
\label{eq:radacc}
\end{equation}
over $N_{\hat{n}} = 36$ directions that uniformly sample all possible vectors in the $\phi=0$ plane. Here, $g_i$ and $\vec{F}_{{\rm UV},i}$ are, respectively, the radiative acceleration and UV flux in direction $i$. Note that we performed a convergence test by doubling the directional resolution to $N_{\hat{n}} = 72$ and find no significant difference in our final results. Indeed, the character of the wind is nearly identical with a change in mass-loss rates within $1\%$ between the two simulations.

The quantity $\mathcal{M}(\mathfrak{t}_i)$ in Equation~\ref{eq:radacc} is the force multiplier, which depends on the ionization state of the material in the cell (see Section~\ref{sec:force_mult}). In the present context, the key point is that the force multiplier is a function of the dimensionless optical depth parameter, $\mathfrak{t}_i$, which is given by 
\begin{equation}
\mathfrak{t}_i=\sigma_e \, \rho \, v_{\rm th} \,  \left|\frac{{\rm d}(\vec{v} \cdot \hat{n}_i)}{{\rm d}s_i}\right|^{-1} .
\label{equation:optcial_depth_param}
\end{equation}
In this expression, ${\rm d}s_i$ represents an infinitesimal step along the direction $\hat{n}_i$, so that $|{\rm d}(\vec{v} \cdot \hat{n}_i)/ {\rm d}s_i| $ is the gradient of the projected velocity component along this direction. Thus $\mathcal{M}(\mathfrak{t}_i)$ is both direction dependent and sensitive to the instantaneous local velocity field.  

These superficially simple expressions capture the complex non-linear coupling between matter and radiation in line-driven flows: the radiation field and ionization state depend on the outflow dynamics (which control the density and velocity fields), but the dynamics also depend on the radiation field and ionization state (which control $F_{{\rm UV},i}$ and $\mathcal{M}$). In our approach, \textsc{pluto} is responsible for updating $\rho$ and $\vec{v}$, while \textsc{python} is responsible for updating (in each cell) $F_{{\rm UV},i}$ and generating a look-up table for $\mathcal{M}(\mathfrak{t}_i)$ across a wide range of $\mathfrak{t}_i$.

In practice, we do not run a \textsc{python} simulation after each hydrodynamic time-step, but only after a fixed interval, $\Delta t_{\rm RAD} \gg \Delta t_{\rm HD}$, of simulation time. This amounts to the assumption that the ionization state of, and radiation field in, each cell are roughly constant over timescales of $\Delta t_{\rm RAD}$. Typically, $\Delta t_{\rm RAD} = 2~\mathrm{s} \simeq 10^3  \Delta t_{\rm HD}$ in our simulations; we have checked that our results are not sensitive to $\Delta t_{\rm RAD}$ in this regime.

After each call to \textsc{python}, we update $\vec{F}_{{\rm UV},i}$ in each cell (see Section~\ref{section:python_method}) and also construct a new look-up table for $\mathcal{M}(\mathfrak{t}_i)$ for the cell, based on its current ionization state (see Section~\ref{sec:force_mult}). However, $\rho$ and $\vec{v}$ in Equation~\ref{equation:optcial_depth_param} {\em are} updated during each hydrodynamic time-step, so the line optical depth parameters, force multipliers and radiative accelerations do evolve on this time-scale as well. 

As noted above, the density at the mid-plane $(\theta=\pi/2)$ is kept fixed and acts
as a mass reservoir. We also impose a a density floor of $\rm{10^{-20}g~cm^{-3}}$
throughout the simulation domain. The mid-plane and inner radial boundary conditions (BCs) 
are both reflective, whilst the $\theta=0$ BC is axisymmetric. An outflow BC is used for 
the outer radial edge of the grid.

\subsection{Ionization and Radiative Transfer}
\label{section:python_method}

Each time \textsc{python} is called after $\Delta t_{\rm RAD}$ of simulation time, it reads in 
the latest snapshot of the density and velocity fields calculated by 
\textsc{pluto}, as well as the wind ionization state produced by 
the previous call to $\textsc{python}$. Photon packets are then injected into the outflow 
by the accretion disc and the wind itself. 
These photon packets are tracked as they make their way through the wind, where they can undergo attenuation 
due to bound-free and free-free opacity, as well as electron scattering and 
bound-bound interactions. 

The photon packets passing through a cell are used to construct estimators that describe the radiation field inside the cell. These are used to update the ionization state of the wind and to calculate the radiative accelerations (see below). 

Unless otherwise indicated, we assume the disc initially has an  effective temperature distribution 
that follows the standard Shakura-Sunyaev (\citeyear{1973A&A....24..337S}) form, 
\begin{equation}
T_{d, {\rm visc}}(R) = T_\star
\left(\frac{R_{\rm WD}}{R}\right)^{3/4}\left(1-\sqrt{\frac{R_{\rm WD}}{R}}\right)^{1/4},
\end{equation}
where $R$ represents the cylindrical radius and 
\begin{equation}
T_{\star} = \left(\frac{3 G M_{\rm WD} \dot{M}_{\rm acc}}{8\sigma\pi R_{\rm WD}^3}\right)^{1/4}.
\end{equation}
Here, $\dot{M}_{\rm acc}$ is the accretion rate through the disc. For the purpose of generating photon packets, the disc is split into concentric annuli, each of which is initially taken to radiate as a blackbody with effective temperature $T_{d, {\rm eff}} = T_{d,{\rm visc}}(R)$. This then fixes both the number and the frequency distribution of the photon packets generated by a given annulus. 

The plasma in the simulation domain itself also produces photons (via free-free, free-bound 
and bound-bound processes). Since the temperature of this material is kept constant in space and time, the 
simulation does not enforce strict energy conservation: the gas in each cell can, in principle, emit more or 
less energy than it absorbs. This is not a serious problem in practice, since typically the plasma is 
not a significant energy source in our simulations. We have nevertheless carried out two tests to check the validity of this approximation. First, as discussed in more detail in Section~\ref{section:discussion}, we have run a much more detailed \textsc{python} simulation for the final state of our fiducial wind model, in which the thermal state of the wind is iterated to convergence along with the ionization state. The resulting 
temperatures are comparable to the $T = 40,000$~K we assume in our RHD simulations. Second, we have repeated 
our fiducial RHD simulation with different gas temperatures ($T = 30,000$~K and $T = 50,000$~K). The 
differences between these runs are relatively minor and do not affect our main results and conclusions. 

 As photons propagate through the wind, they are used to update two sets of estimators related to the radiation field: (i) the angle-averaged mean intensity, $J_\nu$, and (ii) the direction-dependent UV flux, $\vec{F}_{{\rm UV},i}$. As before, the index $i$ here denotes one of $N_{\hat{n}} = 36$ directions chosen to fairly sample all possible vectors in the $\phi=0$ plane. The UV band is defined as running from $7.4\times10^{14}~\mathrm{Hz}$ to $3\times10^{16}~\mathrm{Hz}$. The flux estimator, $\vec{F}_{{\rm UV},i}$, is used by \textsc{pluto} to calculate the radiative accelerations (c.f. Equation~\ref{eq:radacc}), while the mean intensity estimator, $J_\nu$, is used to calculate the ionization state of the gas and the corresponding relationship between the force multiplier and the optical depth parameter (see Section~\ref{sec:force_mult} below). 

Once the photon transport phase is concluded, \textsc{python} calculates the ionization state of the gas based on the local density, temperature and frequency-dependent intensity, $J_\nu$ . The construction of the $J_\nu$ estimator implicitly takes account of any attenuation of photon packets in the wind, as well as any re-emission from the wind itself. Crucially, it also takes 
account of photons arriving in a cell from other parts of the wind, including via scattering events. This scattered and/or reprocessed component of the radiation field makes it much harder for the inner flow to successfully shield material further out. In the context of line-driven disc winds in AGN, it has already been demonstrated that including this component is critically important \citep{2010MNRAS.408.1396S, 2014ApJ...789...19H}. \footnote{Scattered and reprocessed emission plays a much smaller role in our AWD wind models, however, since these outflows are relatively weak and optically thin.}

On each call to \textsc{python}, we carry out at least two iterations of the entire photon generation, radiation transport and ionization equilibrium calculation. This allows us to deal with the effect of irradiation on the disc temperature distribution. Some photon packets generated by the central source / boundary layer -- as well as some photon packets that have undergone reprocessing in the outflow -- will inevitably hit the disc surface. In \textsc{python}, we can treat such photons in one of three ways: (i) we can discard them; (ii) we can assume that they are absorbed and reprocessed; (iii) we can assume that they are specularly reflected. In the simulations described here, we adopt option (ii). Thus, after the first iteration, we update the effective temperature distribution across the disc such that $\sigma T_{d,{\rm eff}}^4(R)= \sigma T_{d,{\rm visc}}^4(R) + F_{\rm irr}(R)$, where $F_{\rm irr}(R)$ is the irradiating flux incident on the disc at radius $R$.

\subsection{Creating Look-Up Tables for the Force Multiplier}
\label{sec:force_mult}

\begin{figure}
\includegraphics[width=\columnwidth]{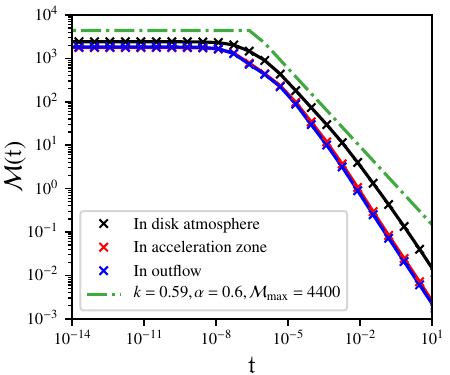}
\caption{The force multiplier $\rm{\mathcal{M}}$ as a function of the dimensionless optical
depth parameter $\mathfrak{t}$ for three representative cells in the simulation domain 
(the locations of the cells are shown in Figure \ref{figure:fiducial_iron_frac} by
colour coded symbols). For comparison, the dot-dashed green line shows the form of $\mathcal{M}(\mathfrak{t})$ for the \citetalias{castor_radiation-driven_1975} $k-\alpha$ approximation (equation~\ref{eq:k-alpha}) adopted by \citetalias{proga_radiation-driven_1998} and \citetalias{dyda_effects_2018} (see section~\ref{section:fiducial} and footnote 4 for an explanation of the value of $k=0.59$).}

\label{figure:m_vs_t}
\end{figure}

After each call to \textsc{python}, we follow the method outlined by \cite{2013ApJ...767..114P} to 
create updated look-up tables for the local force multiplier, based on the SED and ionization parameter 
in each cell. We do not use \textsc{python} directly for this purpose because the calculation of accurate force 
multipliers requires a larger line list than the one used in \textsc{python} (which is sufficient for obtaining the accurate  
estimate of $J_{\nu}$ we need to calculate the ionization state). Instead, we pass the ionization state and the local estimate of $J_\nu$ to a stand-alone code 
that has access to over 450,000 lines  \citep[see][for details]{2013ApJ...767..114P}. For each transition, the code first 
calculates the quantity 
\begin{equation}
    \eta_{u,l}=\frac{hc}{4\pi}\frac{n_lB_{l,u}-n_uB_{u,l}}{\sigma_e\rho v_{\rm th}},
\end{equation}
where $l$ and $u$ refer to the lower and upper level of the relevant transition, respectively. Similarly, $B_{l,u}$ and $B_{u,l}$ are the usual Einstein coefficients for absorption and stimulated emission, respectively. The number 
densities of ions in these levels, $n_l$ and $n_u$, are calculated using the ion densities 
supplied by \textsc{python} and assuming that the excited state populations follow a Boltzmann 
distribution (which is equivalent to assuming that the excited levels are in local thermodynamic 
equilibrium). The force multiplier is then calculated from a weighted sum of contributions from all the lines. In principle, the weighting factor of each transition depends on the specific flux $F_{\nu,i}$ at the wavelength of that transition (see e.g. \citealt{2013ApJ...767..114P}). However, to avoid the computational expense of finding $F_{\nu,i}$ for each direction bin in every grid cell, we make the assumption that this weighting factor can be replaced with its angle-averaged value. That is, we approximate $\frac{F_{\nu,i}}{F_{i}} \approx \frac{J_{\nu}}{J}$, such that
\begin{equation}
    \mathcal{M}(\mathfrak{t})=\sum_{\rm lines}\Delta\nu_D\frac{J_{\nu}}{J}\frac{1-\exp(-\eta_{u,l}\mathfrak{t})}{\mathfrak{t}}
\end{equation}
where $\Delta \nu_D$ is the Doppler width of the line, given by
\begin{equation}
    \Delta \nu_D=\frac{\nu_0 v_{\rm th}}{c},
\end{equation}
where $\nu_0$ is the central frequency of the line.
Figure \ref{figure:m_vs_t} shows the form of $\mathcal{M}(\mathfrak{t})$ for three representative cells in different regions of the simulation, with the form of $\mathcal{M}(\mathfrak{t})$ adopted by \citetalias{proga_radiation-driven_1998} and \citetalias{dyda_effects_2018} shown for comparison. While the shape of the curves is rather similar, the figure shows that the self-consistent calculation of $\mathcal{M}(\mathfrak{t})$ leads to variation of the form of $\mathcal{M}(\mathfrak{t})$ and in general a lower normalisation compared to the idealised \citetalias{castor_radiation-driven_1975} approximation. As we shall see in the next section, these fairly modest differences in $\mathcal{M}(\mathfrak{t})$ lead to real and substantive changes to the character and strength of the resulting outflow in a self-consistent simulation.

\begin{figure*}
\includegraphics[width=1.5\columnwidth]
{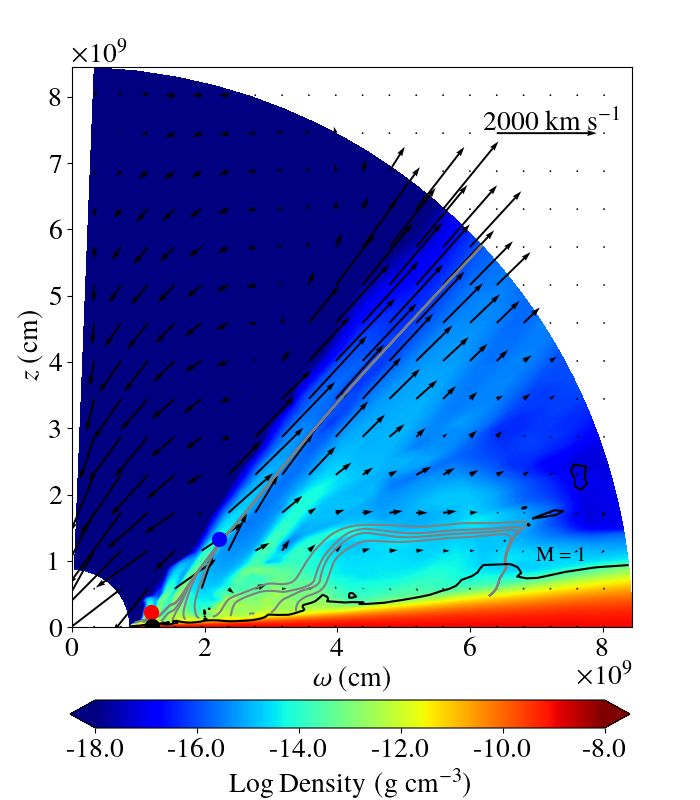}
\caption{The density and velocity structure of the fiducial model. The density is plotted as a colour-map, with velocity vectors overlaid. The grey lines represent streamlines launched
from the $\rm{\theta=86.6\degree}$ surface. The solid black line shows the location of
the Mach~1 surface. The black, red and blue dots show the position of three representative cells in the simulation used for analysis in \autoref{figure:fiducial_iron_frac} and \autoref{figure:fiducial_cell_spec}. See also \autoref{fig:rho_log} in the \S\ref{sec:appendixB} for a version of that figure with logarithmic scales for the axis. A line-driven disc wind is produced, but the flow is significantly weaker and slower than in earlier calculations.}
\label{figure:D_1em9_iso_dens_D}
\end{figure*}

\section{Results}
\label{section:results}

\subsection{Fiducial Model}
\label{section:fiducial}

\begin{figure}
\includegraphics[width=\columnwidth]{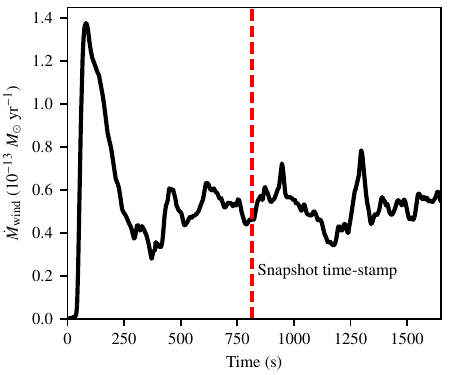}
\caption{The variation of the mass-loss rate through the outer boundary ($\dot{M}_{\rm wind}$) as a function of simulation time in the fiducial model. The vertical dashed line shows the time-stamp used to generate the figures illustrating the model. The high density disc region is excluded in calculating $\dot{M}_{\rm wind}$.}
\label{figure:massloss_vs_time}
\end{figure}

\begin{figure}
\includegraphics[width=\columnwidth]{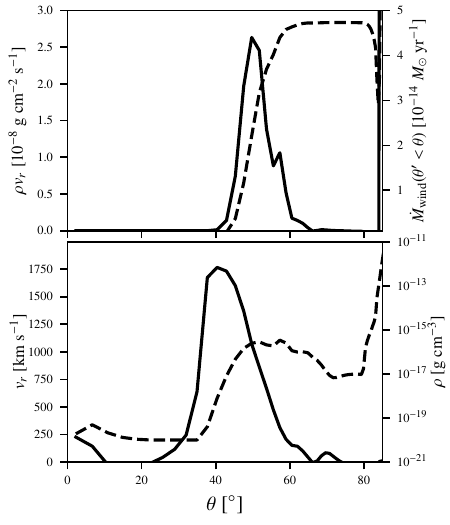}
\caption{Characteristic physical properties of the fiducial model as a function of polar angle, $\theta$.
In the upper panel, we show the product of radial velocity and density, $\rho v_r$, at the outer 
radial boundary (solid line, left scale) and the {\em cumulative} mass-loss
rate, $\dot{M}_{\rm wind} (\theta^\prime < \theta) = \int_0^\theta \frac{{\rm d}\dot{M}_{\rm wind}}{{\rm d}\theta^\prime} {\rm d}\theta^\prime$, through the outer radial boundary (dashed line, right scale). 
The lower panel shows the radial velocity, $v_r$ (solid/left) and density, $\rho$ (dashed/right), again 
measured at the outer radial boundary.}
\label{figure:C_massloss2}
\end{figure}

Figure~\ref{figure:D_1em9_iso_dens_D} shows a snapshot of the density distribution for our fiducial model (HK22D), while 
Figure~\ref{figure:massloss_vs_time} shows the mass-loss rate through the outer boundary of this model as a function 
of simulation time. The density distribution in Figure \ref{figure:D_1em9_iso_dens_D} corresponds to $t = 821$~s of simulation time. This is 
$\simeq$1/5 of the sound crossing time-scale, but Figure~\ref{figure:massloss_vs_time} shows that the flow has converged to a 
quasi-steady state at this point. Figure~\ref{figure:C_massloss2} provides details on the angular distribution of outflow 
properties in the fiducial model, taken from the same snapshot as Figure~\ref{figure:D_1em9_iso_dens_D}. Roughly speaking, 
the main parts of the outflow are concentrated within a 10\degree\ wedge at polar angles between 50\degree\ and 60\degree.

The most important feature of the simulation is the total mass-loss rate, $\dot{M}_{\rm wind} \simeq 5\times10^{-14}~\rm{M_{\odot}{\rm yr}^{-1}}$, i.e. $\dot{M}_{\rm wind} < 10^{-6} \dot{M}_{\rm acc}$. The mass loss rate is roughly 100$\times$ lower than the value seen in the corresponding simulations reported by \citetalias{proga_radiation-driven_1998} and \citetalias{dyda_effects_2018}, which did not include any ionization and radiative transfer calculations. The lower mass-loss rate we find is associated with \emph{both} a lower density in the outflowing wind and a reduced velocity. As discussed in more detail below, this is not just a theoretical issue. As a direct consequence of the reduced mass-loss rate in our simulation, it also fails to reproduce the wind signatures that are commonly observed in AWDs.

Why is the outflow in our simulation so much weaker and slower than in earlier calculations? There are two major contributing factors: 
First, even relatively luminous AWDs -- such as the system described by our fiducial model -- are at best {\em marginally} capable of powering a line-driven wind \citep{2005ASPC..330..103P}. For the physical conditions typically found in such outflows, the {\em maximum} force multiplier that can be achieved is $\mathcal{M}_{\rm max} \simeq 2000$ \citep{1995ApJ...454..410G}. In order to drive strong line-driven disc winds, accreting systems thus require Eddington ratios of at least $\Gamma = L / L_{\rm Edd} \simeq 5 \times 10^{-4}$. As shown in Table~\ref{table:wind_param}, our fiducial model is characterised by $\Gamma = 10^{-3}$, only just above the threshold for efficient line-driving, even under optimal ionization conditions. As a result, successful line-driving in AWDs {\em requires} near-optimal conditions. 

Against this background, the second key factor is that the ionization conditions are {\em not} near-optimal, contrary to what had previously been assumed. \citetalias{proga_radiation-driven_1998} and \citetalias{dyda_effects_2018} adopted the standard \citetalias{castor_radiation-driven_1975} approximation for the force multiplier, $\mathcal{M}(\mathfrak{t}) = k \mathfrak{t}^{-\alpha}$, with parameters based on models of line-driven winds from hot stars, including a hard upper limit of $\mathcal{M}_{\rm max} = 4400$.
With this approximation, $\mathcal{M}(\mathfrak{t}) \simeq \mathcal{M}_{\rm max}$ throughout much of the wind volume, and $\mathcal{M} \gtrsim 1000$ in the acceleration zone (see below and Figure~\ref{figure:force_multipliers}). This allows a strong wind to be driven from the disc. By contrast, our simulations show that the physical conditions in the outflow are actually not all that conducive to line driving, so the system struggles to launch a fast and powerful wind.

The underlying problem is demonstrated quantitatively in Figure~\ref{figure:fiducial_iron_frac}. In hot stars, the single element that contributes most to the line-driving force is Iron \citep[e.g.][]{vinck, noebauer_self-consistent_2015}. There, Iron is mainly found in Fe~\textsc{iv} and Fe~\textsc{v}, and these ionization stages also dominate the  bound-bound opacities. By contrast, the upper left panel of Figure~\ref{figure:fiducial_iron_frac} shows the Iron ionization state 
throughout our fiducial model. The characteristic ionization level is much higher than in hot stars, and the ionization stages required for efficient line driving are largely absent. 

But why is the ionization state so high? In hindsight, this should not come as a surprise, because the maximum temperature in the disc is $T_{d,{\rm max}} \simeq 74,000$~K. Figure~\ref{figure:fiducial_cell_spec} shows the mean intensity in three cells that lie in different regimes of the outflow (the location of these cells is highlighted in Figure~\ref{figure:fiducial_iron_frac}). 
There is significant energy in the SED above the Fe~\textsc{v} ionization threshold, and so the gas is efficiently ionized beyond this stage. 

In order to verify this interpretation, we have tried to replicate the results of \citetalias{proga_radiation-driven_1998} and \citetalias{dyda_effects_2018} by adopting their \citetalias{castor_radiation-driven_1975} approximation for the force multiplier, with $k=0.59$ and $\alpha=0.6$.
\footnote{These parameters are designed to produce equivalent force multipliers to those in \citetalias{proga_radiation-driven_1998} and \citetalias{dyda_effects_2018}. The numerical values for $k$ are different, however, because we use a different value for the thermal velocity.} 
This simulation -- model HK22Df in Table~\ref{table:wind_param} -- is much faster computationally, because there is no ionization step and radiative transfer. \footnote{This means that, as in \citetalias{proga_radiation-driven_1998} and \citetalias{dyda_effects_2018}, the UV flux is not attenuated in the wind, in contrast to our fiducial model with radiative transfer.} Figure~\ref{figure:force_multipliers} allows a comparison of the resulting distribution of the force multiplier to that found in our fiducial model. As expected, $\mathcal{M}$ is much higher when the \citetalias{castor_radiation-driven_1975} approximation is adopted. This over-estimate of the driving force artificially increases the mass-loss rate and wind velocity, bringing them in line with the higher values previously found by \citetalias{proga_radiation-driven_1998} and \citetalias{dyda_effects_2018}. All of this demonstrates how important it is to take full account of ionization and radiation transfer in dynamical simulations of line-driven winds.

\begin{figure*}
\includegraphics{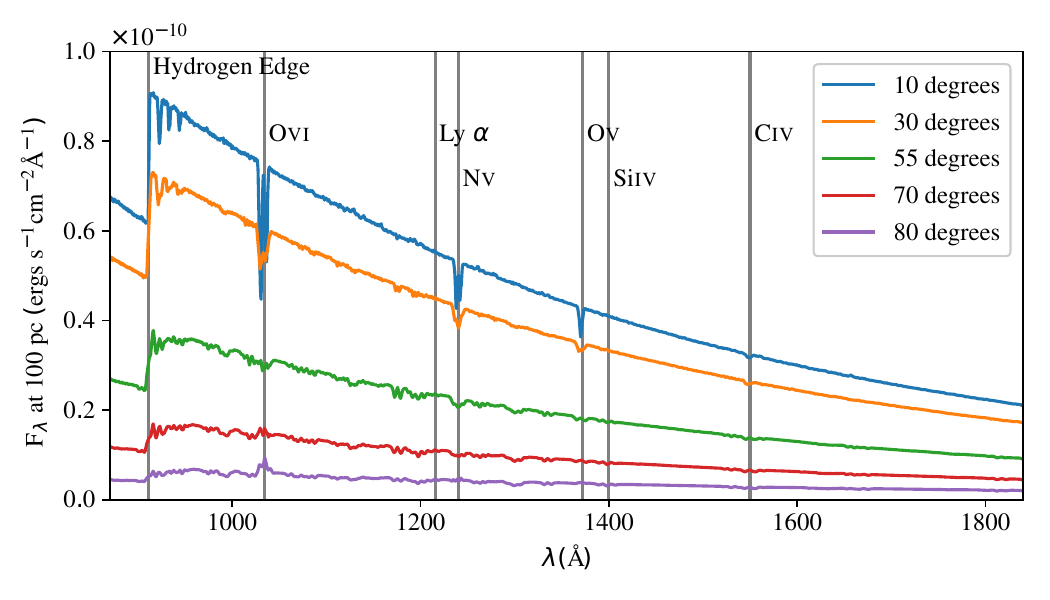}
\caption{Synthetic UV spectra generated from a snapshot of the fiducial model for a range of inclinations. The spectra are normalized such that the flux levels correspond to a system observed at 100~pc. The positions of the Lyman limit and several key UV resonance lines are marked with vertical grey lines.}
\label{figure:fid_spectrum}
\end{figure*}

A major benefit of our RHD approach is that \textsc{python} is also capable of producing  high quality synthetic spectra for any given simulation. Figure~\ref{figure:fid_spectrum} shows a set of UV spectra generated from the snapshot of our fiducial model. There are wind-formed spectral features, but they are all associated with high-ionization transitions. Observationally, the strongest UV lines in wind-driving AWDs tend to be  N~{\sc v}~1240~\AA, Si~{\sc iv}~1400~\AA\ and C~{\sc iv}~1550~\AA. Of these, only N~{\sc v}~1240\AA\ is associated with a reasonably strong feature in the synthetic spectra. Thus, as expected based on the very low mass-loss rate, this simulation is not a good match to observations.

\subsection{Modified Models}
\label{section:modified_models}

Our fiducial model fails to launch a powerful wind because (a) it is barely luminous enough even under optimal conditions, and (b) it is too highly ionized to generate such conditions. Could reasonable changes in the adopted system parameters overcome these challenges? There are two obvious levers. First, we can try to make the system more luminous. The natural ways to do this are to account for the possible contribution of a boundary layer between the WD and the accretion disc or to simply increase the accretion rate. Both of these will, however, produce even harder SEDs, potentially exacerbating the ionization problem. Second, we can try to soften the SED. The only natural way to accomplish this is to reduce the accretion rate, although this will clearly make the system even less luminous. That said, it is well known that disc SEDs are fairly poorly described by sums of blackbodies, and energy dissipation/reprocessing in the upper disc atmosphere and base of the wind might conceivably lead to a softer SED than expected in the Shakura-Sunyaev picture \citep[e.g.][]{k98, shaviv, 2021MNRAS.503.5534H}. This idea can be crudely approximated by adopting a flatter-than-standard effective temperature distribution across the disc. In the rest of this section, we will explore each of these modifications.

\subsubsection{Including a Luminous Boundary Layer}
\label{section:BL}

In the standard Shakura-Sunyaev picture, the accretion disc releases 50\% of the total available accretion luminosity. If the accretor is a non-rotating object with a physical surface (i.e. not a black hole), the rest is radiated away by a narrow boundary layer that connects the surface to the inner disc. At the accretion rates of interest here, the boundary layer in AWDs is expected to be optically thick and radiate approximately as a blackbody. In order to obtain the largest plausible luminosity boost with the smallest plausible effect on ionization, we optimistically assume that the BL covers the entire surface of the WD. This yields the lowest reasonable effective temperature for a luminous BL,  set by $L_{\rm BL} = L_{\rm disc} = 4\pi R^2_{\rm WD} \sigma T^4_{\rm BL}$. For our model parameters, this leads to $T_{\rm BL} \simeq 1.1 \times 10^5$~K.

When we add this boundary layer SED contribution to the simulation (HK22C in Table~\ref{table:wind_param}), its negative effect on the ionization state of the outflow completely dominates its positive effect on the luminosity of the system. In fact, there is no organized fast outflow at all in the final state of this simulation. 

\subsubsection{Increasing the Accretion Rate}
\label{section:hi-Mdot}

Another way to increase the luminosity of the system is to increase the assumed accretion rate. As noted in Section~\ref{section:syspars}, the value 
$\dot{M}_{\rm acc} = \pi \times 10^{-8}~\mathrm{M_{\odot}~{\rm yr}^{-1}}$ adopted in our fiducial model is actually already a high value for AWDs. 
We have nevertheless run a simulation -- HK22F in Table~\ref{table:wind_param} -- in which we increased the accretion rate to $\dot{M}_{\rm acc} = \pi \times 10^{-7}~\mathrm{M_{\odot}~{\rm yr}^{-1}}$. We stress that this is not a realistic accretion rate for ``normal'' AWDs. In fact, it is near or above the threshold where steady nuclear burning would be induced on the WD surface, a situation thought to be found in supersoft X-ray binaries \citep{1992A&A...262...97V}. 

Ignoring these concerns for the moment, we find that this high-$\dot{M}_{\rm acc}$
model does produce a notably stronger line-driven wind, with $\dot{M}_{\rm wind} \simeq 10^{-12}~\mathrm{M_{\odot}~{\rm yr}^{-1}}$. However, this mass-loss rate is still a factor of $\simeq30$ less than obtained by PSD. The characteristic velocity in our simulation, $v \simeq 1000~\mathrm{km~s^{-1}}$, is also much lower than that found by PSD, $v \simeq 7000~\mathrm{km~s^{-1}}$. Finally, the outflow remains very highly ionized and produces no strong UV signatures except O~{\sc vi}~1032~\AA. 

\begin{figure*}
\includegraphics{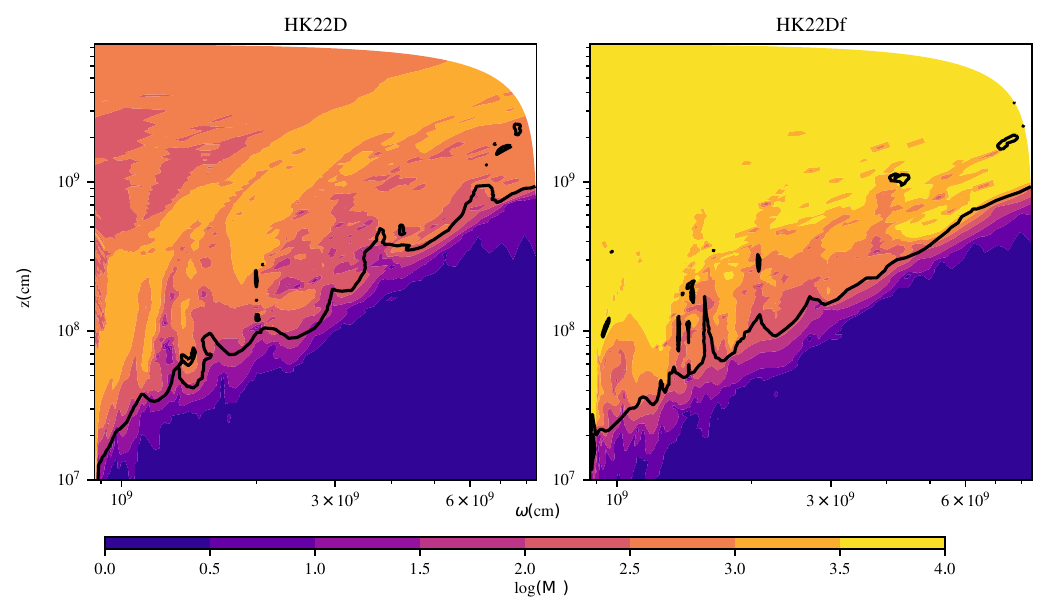}
\caption{The distribution of the force multiplier, $\mathcal{M}$, throughout the fiducial model (left panel, model HK22D) and through an equivalent model with no explicit treatment of ionization and radiative transfer (right panel, model HK22Df; a fixed $\rm{k-\alpha}$ formulation is adopted in this model [see text for details]).  Since $\mathcal{M}$ is direction-dependent, the quantity shown here is the force multiplier in the direction of maximum acceleration. The black lines indicate the Mach~=~1 surfaces.}
\label{figure:force_multipliers}
\end{figure*}

\begin{figure}
\includegraphics[width=\columnwidth,trim={0 2.2cm 0 2cm},clip]{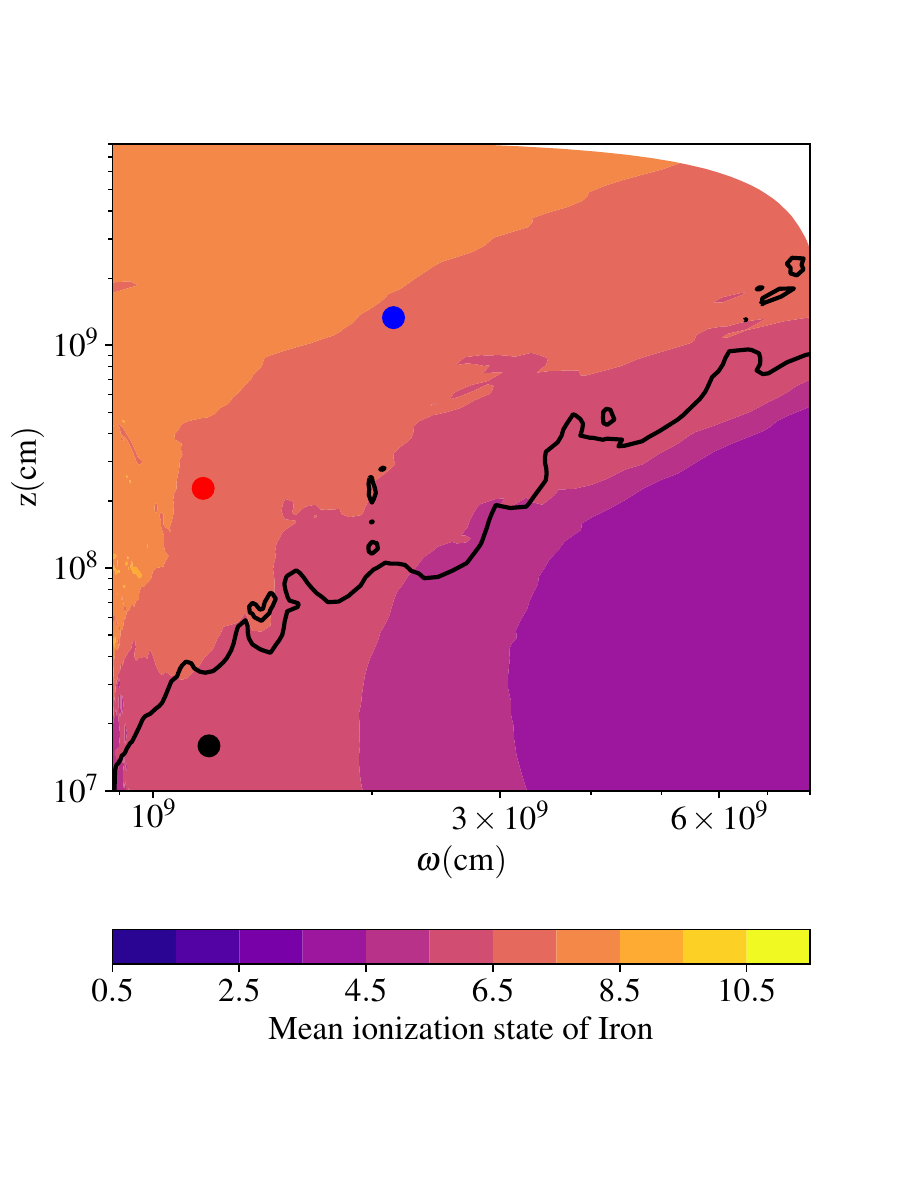}
\caption{The mean ionization state of Iron throughout the fiducial model, using astronomical notation (i.e. neutral~$\equiv 1$). The black, red and blue dots mark the locations of representative cells in the simulation. The 
functional form of the force multiplier, $\mathcal{M}(\mathfrak{t})$, and the mean intensity of the radiation field, $J_\nu$, in these cells are shown in Figures~\ref{figure:force_multipliers} and \ref{figure:fiducial_cell_spec}, respectively. The black line indicates the Mach~=~1 surface.}
\label{figure:fiducial_iron_frac}
\end{figure}

\subsubsection{Decreasing the Accretion Rate}
\label{section:lo-Mdot}

As noted above, we can try to prevent the outflow from becoming over-ionized by lowering the accretion rate, since this will soften the SED. For example, in our fiducial model, the maximum effective temperature in the accretion disc is $T_{d,{\rm max}} \simeq 74,000$~K. Reducing the accretion rate by an order of magnitude, to a value of $\dot{M}_{\rm acc} = \pi \times 10^{-9}~\mathrm{M_{\odot}~{\rm yr}^{-1}}$, lowers this to  $T_{d,{\rm max}} \simeq 41,000$~K. However, this change also reduces the overall luminosity by a factor of 10, bringing the Eddington ratio down to $\Gamma \simeq 10^{-4}$. 

Our RHD simulation for this case is labelled HK22B1 in Table~\ref{table:wind_param}. It does not produce an outflow. This should not come as a surprise, since $\Gamma < \mathcal{M}_{\rm max}^{-1}$ for this model. In fact, the matching outflow in PSD (their Model~1) also failed to produce a supersonic outflow, even though their use of the \citetalias{castor_radiation-driven_1975} approximation assumes near-optimal ionization conditions for line-driving. 

In order to test if there is a ``Goldilocks zone'' for line-driving around $\dot{M}_{\rm acc} \simeq 10^{-8}~\mathrm{M_{\odot}~{\rm yr}^{-1}}$, we have also run a model for this accretion rate. This is listed as HK22B2 in Table~\ref{table:wind_param} and is characterized by $\Gamma \simeq \mathcal{M}_{\rm max}^{-1}$. It, too, does not produce an outflow.

\subsubsection{Flattening the disc Temperature Distribution}
\label{section:flat_Td}

As a final test, we consider a disc that generates the same luminosity as that in our fiducial model, but with a flat effective temperature distribution. For our system parameters, and accounting for the finite radial extent of our simulation domain ($R_{\rm max} = 10~R_{\rm WD}$), the implied constant disc temperature is then $T_{d, {\rm visc}} \simeq 40,000$~K. The corresponding simulation is listed as HK22Ds in Table~\ref{table:wind_param}.

Since the maximum disc temperature is greatly reduced, this model produces a markedly softer SED than our fiducial model, at the same luminosity. Since the ionization state of the wind is now more conducive to line-driving, the mass-loss rate increases by about an order of magnitude to $\dot{M}_{\rm wind} \simeq 4 \times 10^{-13}~\mathrm{M_{\odot}~{\rm yr}^{-1}}$. However, the characteristic wind velocity in this case is also much lower, $v_{\rm wind} \simeq 500~\mathrm{km~s^{-1}}$. This happens because, with fractionally more luminosity emerging at large disc radii, the wind is now driven preferentially from the outer disc. The lower wind speed therefore primarily reflects the reduced Keplerian and dynamical velocities in this region.
\footnote{It should be noted that, because much of the mass loss from the disc takes place near the outer edge of the simulation domain in this model, the characteristic wind velocity and mass-loss rate may be underestimated somewhat.}

\begin{figure}
\includegraphics[width=\columnwidth]{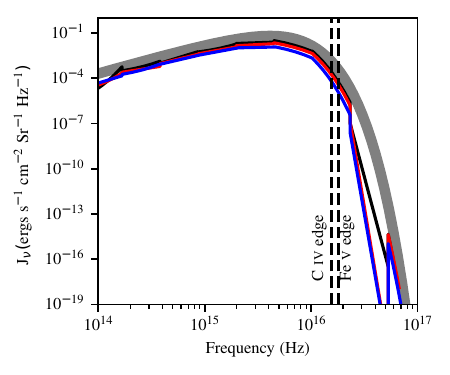}
\caption{The estimator of the mean intensity, $J_\nu$, in three representative cells of the snapshot from our fiducial model. The black, blue and red lines correspond to the corresponding cells  highlighted in Figure \ref{figure:fiducial_iron_frac}. The thick grey line is a 70,000~K blackbody that is shown for reference. The vertical dashed lines mark the important C~{\sc iv} and Fe~{\sc v} ionization edges.}
\label{figure:fiducial_cell_spec}
\end{figure}

\begin{figure*}
\includegraphics{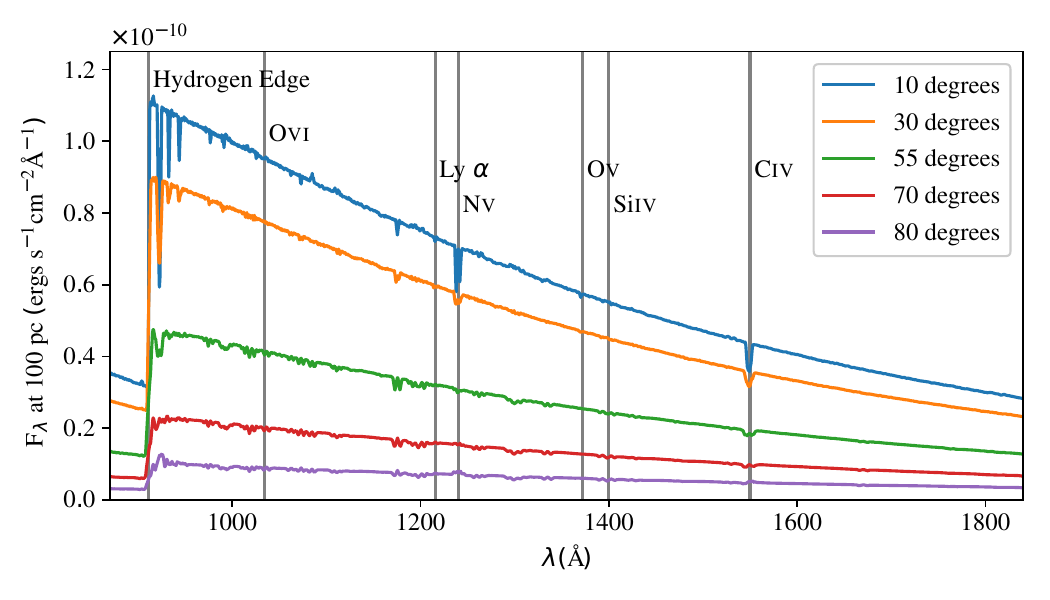}
\caption{Synthetic spectra for model HK22Ds, which is characterised by a constant disc temperature (see discussion in Section~\ref{section:flat_Td}). This may be compared to the same plot for our fiducial model in Figure~\ref{figure:fid_spectrum}. Spectra are shown for a range of inclinations and normalized such that the flux levels correspond to a system observed at 100~pc. The positions of the Lyman limit and several key UV resonance lines are marked with vertical grey lines.}
\label{figure:spectrum2}
\end{figure*}

\begin{figure}
\includegraphics[width=\columnwidth]{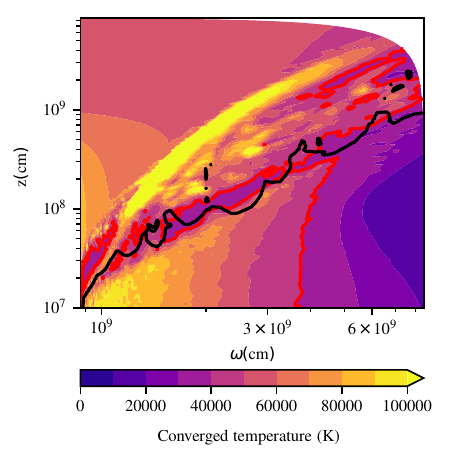}
\caption{The final wind temperatures obtained for a snapshot of the fiducial model after running the stand-alone version of \textsc{python} to convergence on it (such that both ionization/recombination {\em and} heating/cooling are in balance). As usual, the black line marks the Mach~=~1 surface. The red contour marks $T = 40,000$~K, the constant value adopted in our RHD simulations.}
\label{figure:fiducial_conv_temp}
\end{figure}

Figure~\ref{figure:spectrum2} shows the synthetic UV spectra for this model for a range of inclination angles. These spectra do show some of the features seen in observations, notably the N~\textsc{v} and C~\textsc{iv} resonance features at 1240~\AA\ and 1550~\AA, respectively. However, as a result of the low wind velocity in this model, the line profiles are quite narrow and only slightly blue-shifted with respect to their rest wavelengths. There is also no Si~{\sc iv}~1400~\AA\ absorption feature, because the ionization state of the wind is still too high to produce this line. This contrasts with observations of AWDs, in which Si~{\sc iv}~1400~\AA\ is commonly observed \citep[e.g.][]{2002MNRAS.332..127H}.

\section{Discussion and Conclusions}
\label{section:discussion}

We have presented new RHD simulations of line-driven accretion disc winds that include a detailed treatment of ionization and radiative transfer. Focusing on the outflows from AWDs, our main finding is that the physical conditions in these systems are far less conducive to efficient line-driving than implicitly assumed in previous calculations. This is entirely because these outflows turn out to be more highly ionized than those of OB stars. Earlier simulations had adopted approximate force multipliers based on these hot single stars, causing them to systematically overestimate the driving forces.

The overall outflow characteristics are highly sensitive to these issues, because AWDs are only marginally luminous enough to drive a powerful line-driven wind, even under near-optimal ionization conditions. Perhaps most importantly, the mass-loss rates in our simulations are significantly lower than previous estimates, by two orders of magnitude for our fiducial model. In line with this, the corresponding synthetic spectra do not resemble those observed. All of these results raise serious questions about the viability of line-driving as the primary mechanism for generating the winds observed in AWDs.

Fundamentally, our fiducial AWD model struggles to launch a powerful line-driven disc wind for two reasons: 
its relatively low Eddington ratio ($\Gamma = L / L_{\rm Edd} \simeq 10^{-3}$) and the strongly ionizing radiation field produced by its hot disc ($T_{d,{\rm max}} \simeq 74,000$~K).

We have therefore also explored whether simple modifications to the model might generate a stronger outflow that is more in line with observations. 
Including a luminous boundary layer in order to provide more driving photons fails as a strategy: it over-ionizes the outflow even more and causes the outflow to stall completely. Increasing the accretion rate does generate a higher mass-loss rate, but the resulting outflow remains highly ionized and fails to produce the observed wind signatures. This type of model is also physically unrealistic, because accretion at such high rates would likely trigger steady nuclear burning on the WD, making the system a supersoft source, rather than a ``normal'' wind-driving AWD (i.e. a nova-like cataclysmic variable or a dwarf nova in outburst). Softening the SED by {\em reducing} the accretion rate reduces the Eddington ratio even more and again chokes the wind completely. 

The only modification that is marginally successful in both increasing the mass-loss rate and producing more of the observed spectral signatures is to adopt a flat effective temperature profile for the accretion disc. This may be viewed as an ad-hoc attempt to describe the putative reprocessing effect of the disc-wind transition region on the SED. This change leads to an order-of-magnitude increase in $\dot{M}_{\rm wind}$ and the appearance of C~{\sc iv}~1550~\AA\ in the synthetic spectra. However, since this outflow is driven primarily from larger radii, its characteristic speed is significantly lower than implied by observations. The synthetic spectra also presents hardly any wind-formed Si~{\sc iv}~1400~\AA, a relatively low-ionization wind signature commonly seen in AWDs.

Overall, our RHD simulations suggest a rather bleak outlook for line-driving in AWDs: systems luminous enough to drive a wind are also hot enough to over-ionize it. \citet{DP2000} already highlighted the low Eddington ratios found in AWDs as a serious problem for models of line-driven winds in these systems. As also noted by \citet{DP2000}, the only obvious loophole to their conclusion was that the empirically inferred accretion rates (and hence luminosities) of AWDs are systematically underestimated. Our results now close this loophole, since even if AWD were accreting at higher rates than thought they would over-ionize the wind, hence reducing the efficiency of the line-driving mechanism, so as to prevent the wind from becoming massive enough to explain observations. 

Might these conclusions change if we relax one or more of the assumptions and approximations made in our calculations? 
One such assumption is that the accreting material is characterized by solar abundances. The efficiency of line-driving does depend on composition: increasing the abundance(s) of atomic species that contribute significantly to the net line force -- e.g. by increasing the overall metallicity -- will tend to increase the strength of the resulting outflow (\citetalias{castor_radiation-driven_1975}). However, most ``normal'' AWDs appear to exhibit roughly solar or slightly sub-solar abundances \citep[e.g.][]{Harrison2016, Pala2017}. The main exception to this is a sub-set of AWDs in which the accreting material appears to have undergone some degree of CNO processing \citep{Boris2003}. Even in these systems, the line-driving efficiency is unlikely to be significantly different from the solar-abundance case \citep[c.f.][]{Abbott1982}.\footnote{ We note that another distinct class of AWDs may also provide an important test for line-driven disc wind models: AM~CVn stars. In these ultra-short period systems ($P_\mathrm{orb} \lesssim 60~{\mathrm{min}}$), the donor is itself a fully or partially degenerate dwarf. As a result, the accreting material is severely Hydrogen deficient and may show other abundance anomalies also. Interestingly, the prototype of this class displays clear blue-shifted absorption and P-Cygni profiles in its ultraviolet spectrum \citep{Wade2007}.}

Perhaps the most serious remaining approximation is that we treat the wind as isothermal. We are not yet able to include heating and cooling in our models self-consistently, but we expect to be able to in the not too distant future. This will be particularly important for modelling line-driven disc winds in AGN. The SED and geometry of the radiation field are more complex in these systems and may give rise to significant temperature gradients in the outflow. 

In the meantime, we can take a snapshot of our fiducial model and run \textsc{python} on it until thermal equilibrium is achieved. Figure~\ref{figure:fiducial_conv_temp} shows the resulting temperature structure. As expected, the base of the wind heats up to roughly 
the same temperature as the underlying thin disc, and the gas in the acceleration zone settles at temperatures not too far from the $40,000$~K assumed in the RHD simulations. Since the effect of the wind temperature on the ionization state is relatively modest, we believe that relaxing the isothermal approximation in our RHD simulations of AWDs is unlikely to fundamentally change our current conclusions.

A more promising way to make line-driving more efficient in AWDs is to relax the assumption that the disc radiates as an ensemble of blackbodies that follow the Shakura-Sunyaev effective temperature profile. Among all the modifications we have tried so far, our (extremely crude) first attempt to implement this idea -- by imposing a constant effective temperature across the disc -- is the only one that was marginally successful. We have also  carried out some experiments with models in which the disc is represented by an ensemble of stellar atmospheres, rather than an ensemble of blackbodies. Unfortunately, first indications are that the overall disc SED in this case actually has a more pronounced high energy tail, leading to an even \emph{higher} ionization state -- and thus a weaker wind -- than obtained with the blackbody approximation. Nevertheless, if the outflows from AWDs {\em are} due to line-driving, a good description of the disc SED is likely to be the most critical ingredient for future RHD simulations.

Another potentially impactful assumption of our work is that the flow remains smooth on all scales smaller than the grid resolution of our simulations. This may be important because if the flow were clumped on small scales this would effectively increase the density of the material which enhances recombination. Thus small-scale inhomogeneity could help to alleviate the overionization problem. This is a promising avenue for further study since clumping is widely thought to operate in line-driven stellar winds \citep[see e.g.][ and references thererin]{2011IAUS..272..136F} and has also been invoked in disc wind models where it can lead to better agreement with observations \citep[see e.g.][]{2016MNRAS.458..293M}.

There is, of course, no guarantee that line-driving {\em is} responsible for generating the outflows from AWDs. Magnetic fields, in particular, can also drive accretion disc winds, via centrifugal forces \citep[e.g.][]{blandford_hydromagnetic_1982,pelletier_hydromagnetic_1992} and/or magnetic pressure gradients \citep[e.g][]{uchida, stone}.
In fact, there is at least one significant {\em empirical} strike against line-driving in AWDs: multi-epoch UV observations of two wind-driving systems do not exhibit the expected strong correlation between wind activity and continuum level \citep{2002MNRAS.332..127H}. In any case, our simulations clearly show that the nature of the wind-driving mechanism in AWDs is far from settled.

The challenge posed by our simulations for line driving in AWDs is likely to extend to AGN as well. \citet{2010MNRAS.408.1396S} and \citet{2014ApJ...789...19H} already showed explicitly that the line-driven AGN winds in the simulations carried out by PK would, in reality, be over-ionized. In these simulations, the ionization state of the wind was assumed to be dominated by a compact luminous X-ray source that is coincident with an accreting $10^8~{\rm M_{\odot}}$ black hole. Our work here suggests that the same fundamental issue may affect even AGN where the SED is dominated by a relatively cool accretion disc (e.g. X-ray weak quasars with higher black hole masses). The characteristic disc temperatures in such systems are quite similar to those found in luminous AWDs \citep[e.g.][]{james}, so achieving high force multipliers is likely to be difficult. At sufficiently high Eddington ratios, modest force multipliers may be enough to power a radiatively-driven disc wind regardless. However, the viability of line driving across the full spectrum of AGN needs to be urgently reassessed with RHD simulations of the kind we have presented here. 

\section*{Data Availability}
The \textsc{python} and \textsc{pluto} codes used to carry out these simulations are available via the sites \url{https://github.com/agnwinds/python} and \url{http://plutocode.ph.unito.it/} respectively. The data files used to generate the figures presented here are available on request.

\section*{Acknowledgements}
 We would like to thank the anonymous referees for a thorough and helpful report which improved the quality of this work.
We would also like to thank Sergei Dyda for helpful discussions. Calculations in this work made use of the Iridis~5 Supercomputer at the University of Southampton. This work was supported by the UK's Science and Technology Facilities Council [ST/M001326/1, ST/P000198/1]. Partial support for KSL's effort on the project was provided by NASA through grant numbers HST-GO-15984 and HST-GO-16066 from the Space Telescope Science Institute, which is operated by AURA, Inc., under NASA contract NAS 5-26555. JHM acknowledges a Herchel Smith Fellowship at Cambridge. We also gratefully acknowledge the use of the following software packages: matplotlib \citep{Hunter:2007}, astropy \citep{astropy:2018}, {\sc pluto} v4.4 \citep{2007ApJS..170..228M}.

\bibliographystyle{mnras}
\bibliography{bibliography,JM_refs}

\appendix 

\section{Testing our methodology against line-driven OB stellar wind}\label{sec:appendix}

In order to test and validate the method presented in Section \ref{section:method}, we perform a 2D simulation of a OB, line-driven stellar wind and compare our results with the 1D calculations of (\citealt{noebauer_self-consistent_2015}, hereafter NS15). We adopt the same physical parameters as in NS15, i.e. the mass of the central star is $M_\star = 52.5 M_\odot$, the luminosity of the star is $L_\star = 10^6 L_\odot$ and its surface effective temperature (which is also set to be the isothermal temperature of the wind) is $T_\mathrm{eff}=4.2\times 10^4\:K$. The physical domain is extending from 1 to 10 $R_\mathrm{\star}$ in radius, with $R_\mathrm{\star}=1.317\times 10^{12}\:\mathrm{cm}$, and from 0 to $\pi/2$ in latitude. We use the same logarithmically stretched grid as in our disc simulations but with a higher resolution in radius of $N_r=512$ points so as to resolve the sonic point and a slightly lower grid resolution of $N_\theta=32$ to save computational time. We find that the use of a stretched grid in the $\theta$ direction is actually quite important for preserving spherical symmetry throughout the simulation. The ideal strategy is to use a cell ratio in the $\theta$ direction that preserves a constant solid angle across the grid. Otherwise, different level of Monte-Carlo noise between cells can introduce strong asymmetries in the simulation.

The inner radial boundary condition is set to a hydrostatic density profile for the density and so as to conserve the radial flux of mass for the radial velocity when the flow is outgoing (the radial velocity is set to zero otherwise). The outer radial boundary condition is an outflow boundary condition that prevents radial inflow from the outer edge of the simulation. The latitudinal and azimuthal velocities use outflow conditions at both radial boundaries. The latitudinal boundary condition is outflowing for all quantities at the pole and the midplane. 

This stellar wind simulation enables us to test our discretization over direction of the flux (see Section \ref{section:pluto_method}) in a different geometry. It allows us, in particular, to test if our implementation does not introduce any undesired asymmetry in the flow. In this spirit, we use the same number of directions $N_{\hat{n}} = 36$ as in our disc case. Finally, we use $\Delta t_\mathrm{RAD}=200 \Delta t_\mathrm{HD}$ for the coupling between \textsc{PYTHON} and \textsc{PLUTO}.

\begin{figure}
\includegraphics[width=90mm]{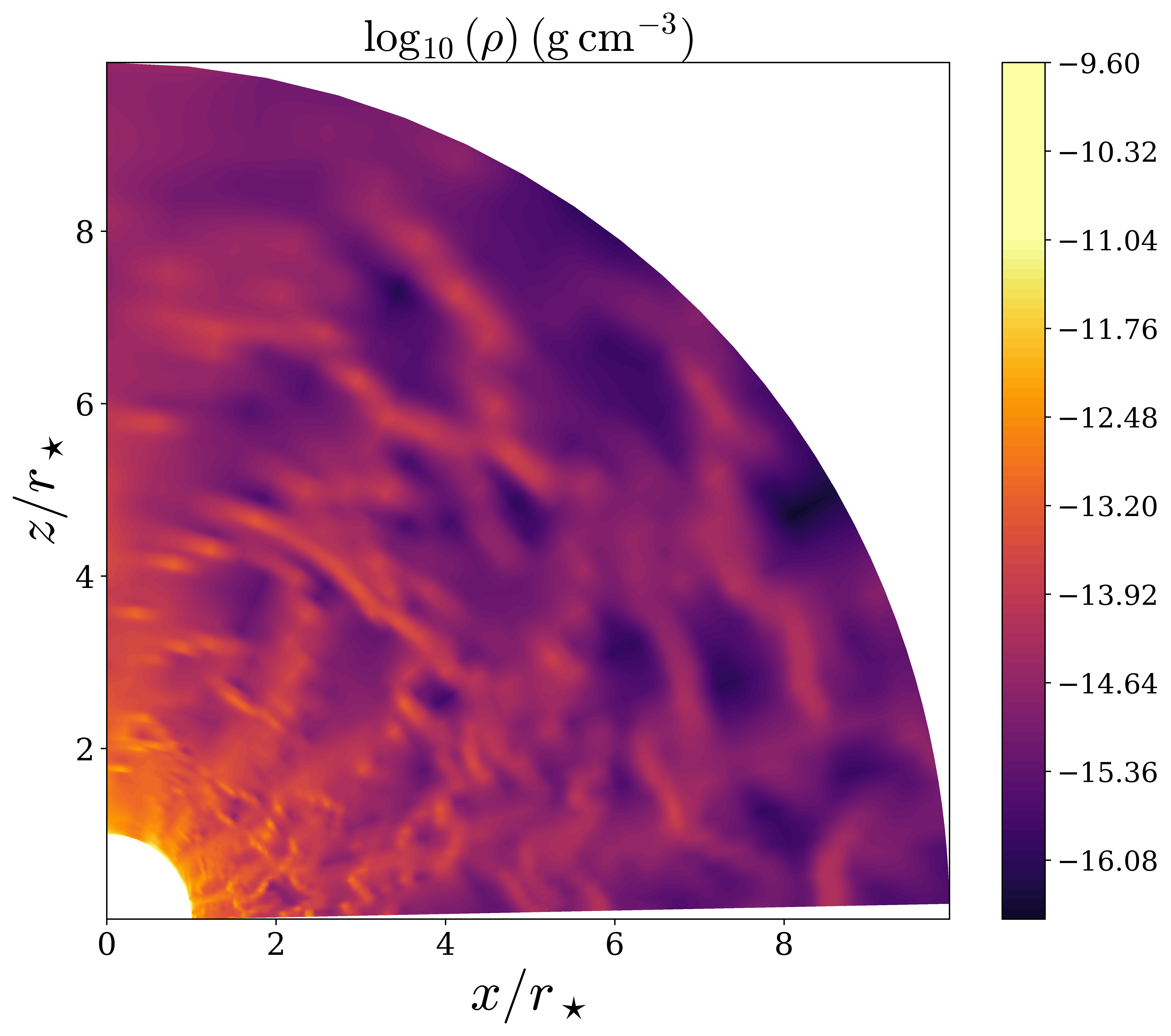}
\caption{2D density map for our stellar wind model. The wind is quite clumpy but we recover a spherically, symmetric wind, validating our 2D implementation of the radiative fluxes and forces presented in \S \ref{section:method}.}
\label{fig:rho_map_sw}
\end{figure}

\begin{figure}
\includegraphics[width=90mm]{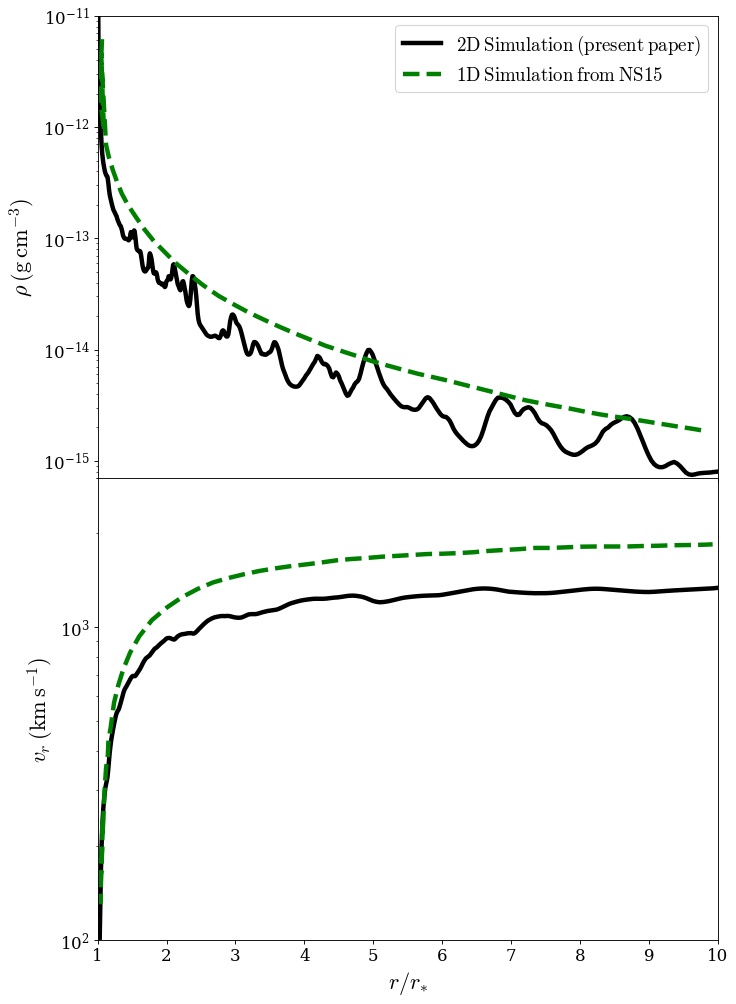}
\caption{Density (top figure) and radial velocity (bottom figure) profiles as a function of radius for our stellar wind model. The solid black line show the (latitudinally-averaged) results of our 2D simulation and the green dashed line show the results from the 1D simulation of NS15 including a finite cone effect and full scattering in the wind. As expected, the results of our 2D simulation are in agreement with the results of NS15. This validates our 2D implementation of the radiative fluxes and forces presented in \S \ref{section:method}.}
\label{fig:rho_vr_sw}
\end{figure}

\autoref{fig:rho_map_sw} shows a snapshot 2D density map from our stellar wind simulation once the mass loss rate has converged to a constant value. The stellar wind is clumpy with density contrasts that can be as large as one order of magnitude. This clumpiness cannot be due to the line-driving instability \citep{owocki_instabilities_1984} since we use the Sobolev approximation in \textsc{python}. Determining whether the clumps are due to a hydrodynamic instability or a radiative/ionization instability is out of the scope of this paper. Despite the clumpiness of the wind, the density structure remains spherically symmetric, with the clumps being distributed isotropically, as expected for a stellar wind. Note that the structure looks slightly more homogeneous near the $z-$pole. This is due to the larger cells in the $\theta$-direction near the pole.

We also plot on \autoref{fig:rho_vr_sw} the latitudinally averaged profiles of the density and the radial velocity as a function of radius. The solid lines show the results of our 2D simulations while the dashed lines show the results from the 1D simulation of NS15, which despite being 1D include a correction for the finite cone effect and the inclusion of scattering in the wind. It is remarkable that despite the presence of large clumps in our simulation, our density and velocity profiles agree quite well (within a factor of 2 for the density and 1.5 for the velocity profiles) with the 1D simulation of NS15. We believe that the quantitative differences between the 1D and 2D simulations could come from the approximated way the finite cone effect is accounted for in 1D and/or the impact of the 2D clumpy structure of the wind.

To conclude, the conservation of the spherical symmetry of the stellar wind (despite its clumpiness) and the agreement between the density and velocity profiles between our 2D simulations and the 1D simulation of NS15 validates the method we presented in Section~\ref{section:method}. The fact that we can reproduce the results of NS15 in hot stellar  winds also demonstrates that the main result of our paper, which is that AWD winds are overionized and so cannot reproduce the observed mass-loss rate, is not due to our method but due to improved self-consistent treatment of the interaction between matter and radiation compared to previous studies.

\section{Logarithmic density map}\label{sec:appendixB}

\begin{figure}
\includegraphics[width=100mm]{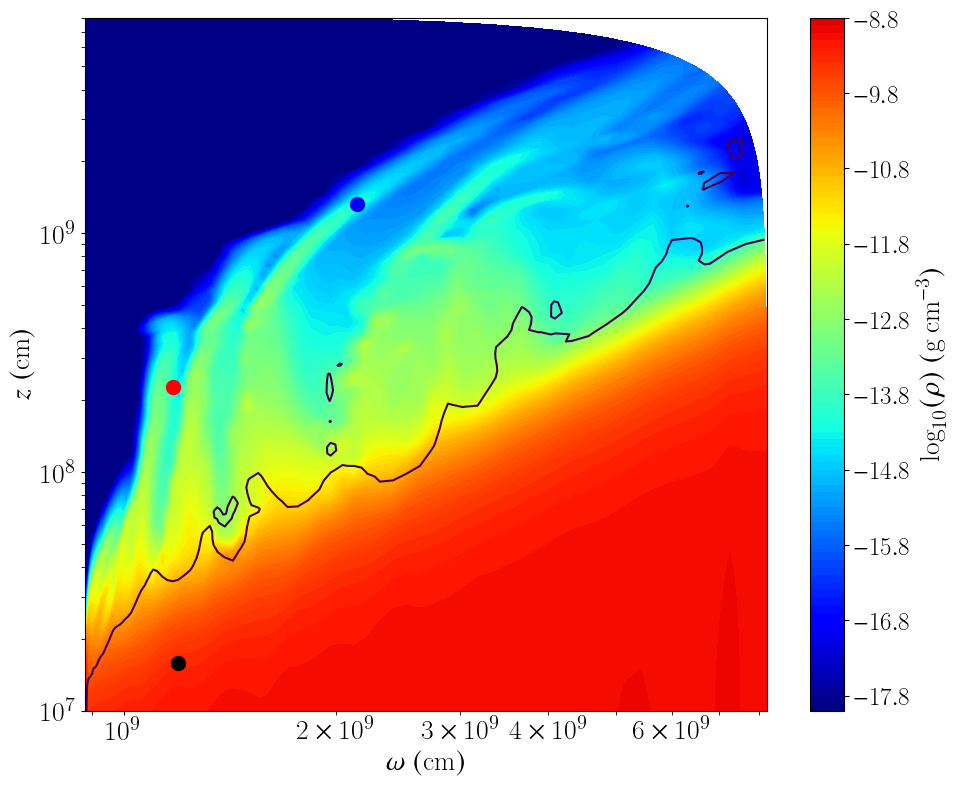}
\caption{Density map for our fiducial model with logarithmic axes. The solid black line shows the location of the Mach 1 surface. The black, red and blue dots show the position of three representative cells used for analysis in \autoref{figure:fiducial_iron_frac} and \autoref{figure:fiducial_cell_spec}.}
\label{fig:rho_log}
\end{figure}

In order to facilitate comparisons between \autoref{figure:force_multipliers}, \autoref{figure:fiducial_iron_frac}, \autoref{figure:fiducial_conv_temp} and the density structure shown in \autoref{figure:D_1em9_iso_dens_D}, we provide in 
\autoref{fig:rho_log} a map of the density with the axes on a logarithmic scale.

\label{lastpage}

\bsp	

\end{document}